\documentclass[pra,aps,twocolumn,showpacs]{revtex4}
\usepackage{amsfonts,amsmath,bm}

\usepackage{graphicx}

\begin{document}

\title{Collective spontaneous emission from a system of quantum dots} 
\author{Wildan Abdussalam}
\author{Pawe{\l} Machnikowski}
 \email{Pawel.Machnikowski@pwr.wroc.pl}  
\affiliation{Institute of Physics, Wroc{\l}aw University of
Technology, 50-370 Wroc{\l}aw, Poland}

\begin{abstract}
We study the spontaneous emission from a regular lateral array or a
randomly distributed ensemble of quantum dots under strong excitation
(full inversion) conditions. We focus on the similarities and
differences between the cases of random and regular arrangement of the
dots and show that there is very little difference between the
evolution of luminescence in these two cases, both for identical dots
and for a realistically inhomogeneously broadened ensemble. This means
that the enhanced emission or superradiance effect is not due to
accidental clustering of pairs of dots. Moreover, we point out that
observation of an enhanced emission under weak excitation does not
prove that true superradiance will develop in a fully inverted system.
\end{abstract}


\maketitle

\section{Introduction} 
\label{sec:intro}

The phenomenon of collectively enhanced emission, 
described by Dicke for ensembles of
atoms \cite{Dicke}, can also occur in quantum dots (QDs)
\cite{scheibner07}. The cooperative radiation effect is
due to the collective interaction of the two
emitters with quantum radiation field. In the case of
non-identical dots and in the absence of
interactions, the collective emission appears if the
interband transition energies in various QDs differ by no more than the
emission line width  \cite{sitek07a}, which requires the dots to be nearly identical, beyond
the recent technological possibilities. However, coupling between the
dots restores the collective nature of the emission and leads to
accelerated or slowed down emission even for dots with different
transitions energies \cite{sitek09ab}.  

Two typical couplings that may appear in a system of QDs are long-range 
Coulomb interactions and short-range couplings that may result from a
combination of carrier tunneling and Coulomb correlation effects. As we
have shown, these two
couplings show no qualitative difference in either accelerating or
slowing down the emission in spite of their essentially different
physical origin and properties \cite{Wil}. The simulations of
spontaneous emission from ensembles of coupled QDs show that the
emission rate in such a system can indeed be increased due to
collective coupling of the emitters to the electromagnetic field
\cite{kozub}. As the positions of the dots are random within
the sample plane, it might happen that this enhanced emission
effect results from clustering of dots in a
self-assembled sample, as a result of which strongly coupled pairs of
dots are formed and contribute to the emission. Therefore, in order to
fully understand the role of the 
inter-dot coupling it is interesting to study the role of the planar
arrangement of the dots on the observed collective emission.
Moreover, so far the collective emission from QD ensembles was
discussed only under weak excitation conditions (one excitation
delocalized over the ensemble) \cite{kozub}, where it is
manifested only in the form of an enhanced emission rate. 
In order to observe actual superradiance, that is, a
non-monotonic evolution of the emission with a delayed outburst of
radiation \cite{Skri}, one has to study the photon
emission under strong excitation conditions. 

In this paper, we study the collective spontaneous emission from small
ensembles of QDs under strong excitation (full inversion)
conditions for two kinds of spatial arrangements of the QDs in the
sample plane: regular arrays and randomly scattered dots. We show
that the dynamics of emission very weakly depends on the QD
arrangement. 

The paper is organized as follows. In Sec.~\ref{sec:model}, we
describe the model of a small ensemble of QDs in either random or
regular planar arrangement. Next, in
Sec.~\ref{sec:results},  we present and discuss the simulation results for the
photon emission in these two cases. Finally, Sec.~\ref{sec:concl} 
concludes the paper. 

\section{Model} 
\label{sec:model}

\begin{figure}[tb]
\includegraphics[width=85mm]{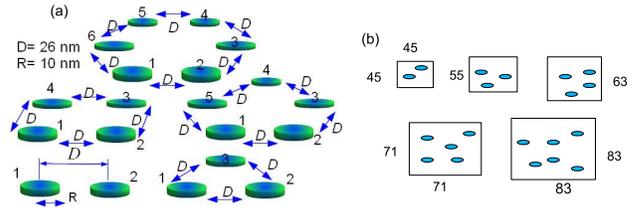}
\caption{\label{fig:diagram}The planar arrangement of the QDs: (a) regular array; (b)
 random ensemble (sample dimensions in nanometers).} 
\end{figure}

We consider an ensemble of a few (up to six) QDs placed in the $xy$ plane at
positions $\bm{r}_{\alpha}$, where $\alpha$ numbers the dots. Each QD
is modeled as a two-level system (empty  
dot and one exciton) with the fundamental transition energy
$E_{\alpha}=\overline{E}+\epsilon_{\alpha}$, where $\overline{E}$ is
the average transition energy in the ensemble. 
We consider two different arrangements of the QDs in the sample plane:
Regular arrays of QDs (double dots, triangular, rectangular
aligned structure of the dots, penta- and hexagonal structures, see
Fig.~\ref{fig:diagram}(a)) and ensembles of randomly distributed QDs, as
schematically depicted in Fig.~\ref{fig:diagram}(b). 
In the latter case, the ensemble is
randomly distributed with a fixed surface density over square mesas
with the restriction that the 
inter-dot distance cannot be smaller than 10~nm. The
spectral properties of the dots are modeled 
by a Gaussian distribution of their transition energies $E_{\alpha}$
with the standard deviation $\sigma$ around the mean $\overline{E}$. 
The dots are coupled by an interaction $V_{\alpha\beta}$ which can be either
of dipole--dipole character (long-range \textit{dispersion force}) or
short-range (exponentially decaying with the distance).

We introduce the transition
operators for the dots: the ``exciton annihilation'' operators
$\sigma_{\alpha}$ which annihilate an exciton in the dot $\alpha$,  
and the ``'exciton creation'' operators $\sigma_{\alpha}^{\dagger}$
which creates an exciton in the dot $\alpha$.  
The exciton number operator for the dot $\alpha$ is then
$\hat{n}_{\alpha}=\sigma_{\alpha}^{\dag}\sigma_{\alpha}$. 
Using these operators, the Hamiltonian of the many dots system is
written in the frame rotating with the frequency $\overline{E}/\hbar$
in the form  
\begin{displaymath}
\begin{array}{lll}
H = \sum_{\alpha =1}^{N} \epsilon_\alpha \sigma_\alpha^\dagger \sigma_\alpha + \sum_{\alpha \neq \beta = 1}^N V_{\alpha \beta} \sigma_{\alpha}^{\dagger} \sigma_{\beta}.
\end{array}
\end{displaymath}

The long-range dipole coupling is described by
\begin{displaymath}
V_{\alpha\beta}=V^{(\mathrm{lr})}_{\alpha\beta}
=-\hbar\Gamma_{0}G(k_{0}r_{\alpha\beta}),
\end{displaymath}
where $\bm{r}_{\alpha\beta}=\bm{r}_{\alpha}-\bm{r}_{\beta}$,
\begin{displaymath}
\Gamma_{0}=
\frac{|d_{0}|^{2}k_{0}^{3}}{3\pi\varepsilon_{0}\varepsilon_{\mathrm{r}}},
\end{displaymath}
is the spontaneous emission (radiative recombination) rate for a
single dot, 
$\varepsilon_{0}$ is the vacuum permittivity,
$\varepsilon_{\mathrm{r}}$ is the relative dielectric constant of the
semiconductor, $k_{0}=n\overline{E}/(\hbar c)$,
where $c$ is the speed of light and
$n=\sqrt{\varepsilon_{\mathrm{r}}}$ is the refractive index of the
semiconductor, 
and
\begin{eqnarray*}
G(x) & = & \frac{3}{4}\left[
-\left( 1-|\hat{\bm{d}}\cdot\hat{\bm{r}}_{\alpha\beta}|^{2}  \right)
\frac{\cos x}{x} \right. \\
&& \left.+\left( 1-3|\hat{\bm{d}}\cdot\hat{\bm{r}}_{\alpha\beta}|^{2}  \right)
\left( \frac{\sin x}{x^{2}}+\frac{\cos x}{x^{3}} \right)
 \right],
\end{eqnarray*}
where
$\hat{\bm{r}}_{\alpha\beta}=\bm{r}_{\alpha\beta}/r_{\alpha\beta}$ and
$\hat{\bm{d}}=\bm{d}/d$, where $\bm{d}$ is the interband matrix
element of the dipole moment operator which is assumed identical for
both dots. For a heavy hole exciton,
$\bm{d}=(d_{0}/\sqrt{2}) [1 , \pm i, 0]^{T}$, so that for a vector
$\bm{r}_{\alpha\beta}$ in the $xy$ plane one has
$|\hat{\bm{d}}\cdot\hat{\bm{r}}_{\alpha\beta}|^{2}=1/2$.
The short-range coupling is described by 
\begin{displaymath}
V_{\alpha\beta}=V^{(\mathrm{sr})}_{\alpha\beta}
=V_{0}e^{-r_{\alpha\beta}/r_{0}}.
\end{displaymath}

\begin{figure}[tb]
\includegraphics[width=85mm]{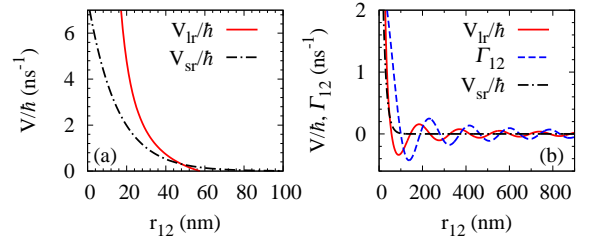}
\caption{\label{fig:GV} The interference term of the decay rate $\Gamma_{12}$ and
  the short- and long-range coupling amplitudes
  $V_{\mathrm{lr}},V_{\mathrm{sr}}$ as a function of the inter-dot
  distance. In (a), the small distance section is shown, while in (b)
  the oscillating tail at larger distances is visible.}
\end{figure}

The effect of the coupling to the radiation field is
accounted for by including the dissipative term in the evolution equations,
which describes radiative recombination of excitons.
The equation of evolution of the density matrix is then given
by \cite{lehmberg70a} 
\begin{equation}
\label{evol}
\dot{\rho}=-\frac{i}{\hbar}[H_{0},\rho]+
\sum_{\alpha,\beta=1}^{N}\Gamma_{\alpha\beta}\left[ 
\sigma_{\alpha}\rho\sigma_{\beta}^{\dag}
-\frac{1}{2}\left\{ \sigma_{\alpha}^{\dag}\sigma_{\beta},\rho \right\}_{+}
 \right], 
\end{equation}
where $\Gamma_{\alpha\alpha}=\Gamma_{\beta\beta}=\Gamma_{0},\quad
\Gamma_{\alpha\beta}=\Gamma_{\beta\alpha}=\Gamma_{0}F(k_{0}r_{\alpha\beta})$,
with 
\begin{eqnarray*}
F(x) & = & \frac{3}{2}\left[
\left( 1-|\hat{\bm{d}}\cdot\hat{\bm{r}}_{\alpha\beta}|^{2}  \right)
\frac{\sin x}{x} \right.\\
&&\left.+\left( 1-3|\hat{\bm{d}}\cdot\hat{\bm{r}}_{\alpha\beta}|^{2}  \right)
\left( \frac{\cos x}{x^{2}}-\frac{\sin x}{x^{3}} \right)
 \right],
\end{eqnarray*}
and $\{\ldots,\ldots\}_{+}$ denotes the anticommutator. The diagonal
decay rates $\Gamma_{\alpha\alpha}$ describe the emission properties
from a single dot, while the off-diagonal terms
$\Gamma_{\alpha\beta}$, $\alpha\neq\beta$, account for the interference
of emission amplitudes resulting from the interaction with a common
reservoir and are responsible for the collective effects in the emission.
The values of the two couplings as well as the interference term of
the decay rate  $\Gamma_{\alpha\beta}$ are plotted as a function of the
distance between the dots in Fig.~\ref{fig:GV}.

In our simulations, we  use the parameters for a typical CdSe/ZnSe QD system:
$\Gamma_{0}=2.56$~ns$^{-1}$, $n=2.6$, the average transition energy of the QD ensemble $\overline{E}=2.59$~eV and the QD surface density $\nu=10^{11}$ $QDs/cm^2$. For the
tunnel coupling we choose the amplitude $V_{0}=5$~meV and the range
$r_{0}=15$~nm.

\section{Results}
\label{sec:results}

\begin{figure}[tb]
\includegraphics[width=85mm]{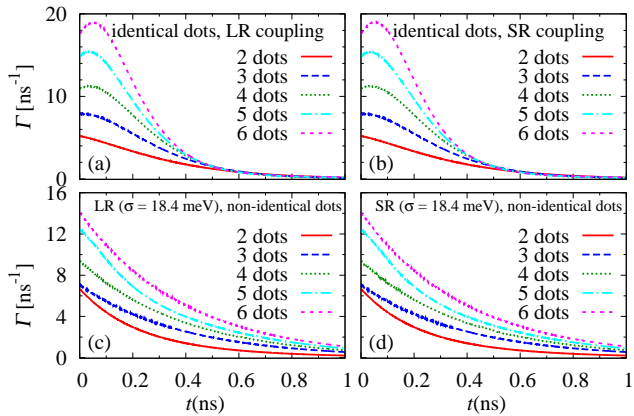}
\caption{\label{fig:g14}The evolution of photon emission for the
  fully inverted  initial state in regular arrays of a few QDs.}
\end{figure}

In Fig. \ref{fig:g14} we show the results of numerical simulation
based on Eq. \ref{evol} for regular QD arrays. 
At the initial time, the system is strongly excited to the fully
inverted state
$\vert\psi(0)\rangle$ = $\vert 1..1..1\rangle$ and we subsequently
determine the photon emission rate as a function of time.
On each plot, the photon emission in the
system is shown as a function of time for identical dots ($\sigma =
0$) and for non-identical dots with the realistic value of the
fundamental transition energy standard deviation $\sigma = 18.4$~meV

As can be seen in Figs.~\ref{fig:g14}(a) and (b), the spontaneous
emission from a regular array of identical QDs develops a maximum,
which is a few-emitter counterpart of the 
delayed outburst of photon emission observed in macroscopic samples
\cite{Skri}. This effect appears already for three QDs and becomes
more and more pronounced as the number of dots increases. It implies
that the true superradiance effect appears in the regularly arranged 
small ensemble of identical QDs with either long- or short-range
couplings. However, if the array is formed by non-identical dots with a realistic 
inhomogeneity of the fundamental transition energies
(Fig.~\ref{fig:g14}(c) and (d)), when the value of $\sigma$ 
is greater than both the coupling strengths and the inverse exciton
life time (the emission line width), then the maximum is completely
washed out. Thus, under conditions that led to enhanced emission in
the weak excitation regime \cite{kozub}, full superradiance effects
under strong excitation are not observed.

\begin{figure}[tb]
\begin{center}
\includegraphics[width=85mm]{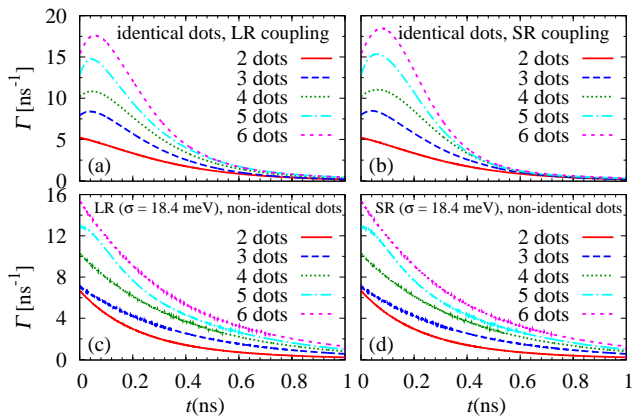}
\end{center}
\caption{\label{fig:g17}The evolution of photon emission for the
  fully inverted  initial state in an ensemble of a few QDs with a
  random spatial distribution.}
\end{figure}
 
In order to find out whether a randomly distributed ensemble of QDs
behaves similar to a regular array of QDs we now study the photon
emission from a random QD ensemble in the same strong excitation
regime.
In each case, we calculate an average of
100 realizations of the evolution in systems with a given
number of dots but differing in their positions within the sample
plane.
Figs. \ref{fig:g17}(a) and (b) show the simulated photon emission for
an ensemble of identical dots. Again, a superradiant maximum builds up
as the number of QDs grows. Similarly to the case of regular QD
arrays, the maximum disappears and the decay of luminescence becomes
monotonic when the realistic distribution of fundamental transition
energies is taken into account (Figs. \ref{fig:g17}(c) and (d)). 

By comparing Fig.~\ref{fig:g14} with Fig.~\ref{fig:g17} one can see
that there is very little difference in arranging the dots either
randomly or regularly under strong excitation
conditions. Regular ordering of QDs only slightly enhances the peak
emission in the hypothetical case of identical dots and slightly
suppresses the emission in the realistic case of an inhomogeneously
broadened ensemble.

\section{Conclusions}
\label{sec:concl}

We have shown that the delayed outburst of radiation, typical for the
superradiant emission develops in the luminescence from regular arrays or
randomly distributed ensembles of quantum dots in the strong
excitation regime.  
The way the dots are distributed in the sample plane (regular
vs. random) makes very little difference on the photon emission. 
This means that the system response is not dominated by
accidental clustering that might appear in the random distribution
case and lead to strongly enhanced contribution to the overall
emission  from pairs of accidentally very closely spaced dots.
Moreover, in both cases, the superradiant maximum is completely washed out if
the realistic degree of inhomogeneity of the fundamental transition
energies is taken into account. 
Thus, enhanced emission observed experimentally under weak
excitation does not imply that true superradiance will be manifested
for a fully inverted system.

\textbf{Acknowledgment:} This work was supported by the Foundation for
Polish Science under the TEAM programme, co-financed by the European
Regional Development Fund.


\end{document}